\documentclass[prl,aps,twocolumn,showpacs,amsmath,amssymb,amsfonts,floatfix,superscriptaddress]{revtex4}
\usepackage{graphicx}
\usepackage{dcolumn}
\usepackage{bm}
\usepackage{latexsym}
\usepackage{float}
\usepackage{supertabular}
\usepackage{longtable}
\usepackage{hyperref}

\usepackage{color} 
\usepackage{verbatim}
\usepackage{comment,xcolor}

\begin{document}
\title{Evolution of a Fluctuating Population in a Randomly Switching Environment}
\author{Karl Wienand}
\author{Erwin Frey}
\affiliation{Arnold Sommerfeld Center for Theoretical Physics, Department of Physics, 
Ludwig-Maximilians-Universit\"at M\"unchen, Theresienstrasse 37, 80333 M\"unchen, Germany}
\author{Mauro Mobilia}
\affiliation{Department of Applied Mathematics, School of Mathematics, University of Leeds, Leeds LS2 9JT, U.K.}
\email{M.Mobilia@leeds.ac.uk}
\begin{abstract}
Environment plays a fundamental role in the competition for resources, and hence in the evolution of populations.
Here, we study a well-mixed, finite population consisting of two strains competing for the limited resources
provided by  an environment that randomly switches  between states of  abundance and scarcity.
Assuming that one strain grows slightly faster than the other, we consider 
two scenarios---one of pure resource competition, and one in which one strain provides a public good---and 
investigate how environmental randomness (external noise)  {\it coupled} to demographic (internal) noise 
determines the population's fixation properties and  size distribution.
By analytical means and simulations, we show that these coupled sources of noise can 
significantly enhance the fixation probability of the slower-growing species.
We also show that the population size distribution can be unimodal, bimodal or multimodal
and undergoes  noise-induced transitions between these regimes when the rate of switching matches the
population's growth rate.
\end{abstract}

\maketitle
\begin{figure}[htb]
\includegraphics{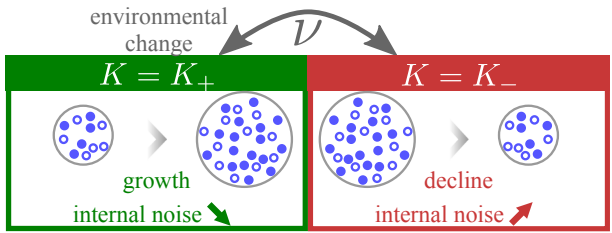}
\caption{{\it (Color online)}. Cartoon of the model: 
Coupled evolution of the population size and its composition, consisting of strains $S$ ($\circ$) and $F$ ($\bullet$), subject to
a stochastically switching carrying capacity $K(t)\in \{K_-,K_+\}$, see Eq.~(\ref{eq:K(t)}).
$K$ switches with rate $\nu$ from $K_-$ to $K_+$, leading to population growth and decreasing demographic fluctuations (internal noise).
When $K$ switches (with rate $\nu$) from $K_+$ to $K_-$, the population size declines and demographic fluctuations increase.}
\label{fig:Cartoon}
\end{figure}

Natural populations face ever-changing environmental conditions, which influence their evolutionary fate.
For instance, the abundance of nutrients, the presence of toxins, or external factors like temperature and pH often 
influence the evolution of species~\cite{Morley83,Fux05}.
Several mechanisms have been suggested for a population to cope with fluctuating environments, such as 
phenotypic  heterogeneity, bet-hedging, and storing the gains realized during good periods~\cite{Chesson81,Kussell05,Acar08,Beaumont09,Visco10}.
The impact of random environmental changes (external noise) on fitness variability has been studied in population 
genetics, predator-prey systems, as well as in game-theoretic and related 
models~\cite{May73,Karlin74,He10,Tauber13,Danino16,Assaf12,AMR13,Ashcroft14,Melbinger15,Hufton16,Hidalgo17,Kussell05b}.
Demographic fluctuations (internal noise), arising in finite populations, 
are responsible for fixation---when one species takes over the population~\cite{Kimura,Ewens},
and  determine the population's internal composition. Internal noise is
stronger in small populations and becomes negligible in large ones.
The dynamics of the population composition is often coupled with the evolution  
of its size~\cite{Roughgarden79,Leibler09,Melbinger2010,Cremer2011,Melbinger2015a}.
This may result in a coupling of environmental and internal noise, with
external randomness affecting the population size which in turn modulates demographic fluctuations.
The interdependence of external and internal noise is especially relevant to  microbial communities, 
which can experience sudden, extreme environmental changes~\cite{Wahl02,Patwas09,Wienand15,Brockhurst07a,Brockhurst07b}.
These may lead to \textit{population bottlenecks}: new colonies or biofilms formed from only few individuals, thus prone to fluctuations.
This mechanism leads to feedback loops between social interactions and environment, and to population dynamics of great evolutionary relevance~\cite{Wahl02,Patwas09,Wienand15}.
For instance, recent experiments on {\it Pseudomonas fluorescens} showed  that the formation and sudden 
collapse of biofilms promotes the evolution of cooperative behaviors~\cite{Brockhurst07a,Brockhurst07b}.
\\
Most studies, however, treat environmental and internal noise 
independently~\cite{May73,Karlin74,He10,Tauber13,Danino16,Assaf12,AMR13,Ashcroft14,Kussell05b,Melbinger15,Hufton16,Hidalgo17}. Moreover, 
 environmental randomness is often modeled with white noise~\cite{May73,Karlin74,Melbinger15}, 
although the correlation time is finite  in realistic settings.
Here, we develop an approach to study the {\it coupled effect} of environmental and 
internal noise on the evolution of a two-species population in a stochastic environment:
We assume that the  carrying capacity randomly switches between two values, following a 
 dichotomous noise~\cite{HL06,Bena06}.
A distinctive feature of this model is the coupling of internal and environmental noise (Fig. \ref{fig:Cartoon}): 
Demographic fluctuations depend on the population size which varies  following the switching environment.
We first consider a scenario with pure resource competition, in which 
the dynamics of the population  composition and its size are only linked by demographic fluctuations.
Then, we investigate a public good scenario in which interspecies social interactions 
explicitly couple the composition and ecological (size) dynamics.
Using analytical and computational means, we show how environmental and internal noise can significantly influence 
the population's fixation properties.
Moreover, we show that external noise induces a transition between different regimes
of the population size distribution.

We consider a well-mixed population of finite and time-fluctuating size $N(t)=N_{S}(t) + N_{F}(t)$
consisting of two strains. At time $t$,  $N_S(t)$ individuals are of a slow-growing strain $S$, 
corresponding to a fraction $x=N_S/N$ of the population, 
and $N_F$ are of a fast-growing species $F$.
Individuals of strain $\alpha\in\{S,F\}$ reproduce  with a per-capita rate $T_\alpha^+=f_\alpha/\bar 
f$~\cite{Melbinger2010,Cremer2011}, where $f_\alpha$ is the fitness of strain $\alpha$ and $\bar f=x f_S + (1-x)f_F$ is 
the average fitness.
Here $f_F=1$ and $f_S=1-s$, where $0<s\ll1$ denotes the 
weak selection intensity that disadvantages the strain $S$~\cite{Kimura}.
The population size growth often depends on its composition, {\it e.g.}  one strain may produce a public good.
This is accounted for by multiplying the birth rates $T_\alpha^+$ by a ``global fitness'' 
$g(x)$~\cite{Roughgarden79,Melbinger2010,Cremer2011}.
Here, we focus on two important cases:
(i) {\it pure resource competition}: $g(x)=1$, in this setting $x$ and $N$ are only coupled by fluctuations; and
(ii) {\it public good}: $g(x)=1+bx$, corresponding to an explicit coupling of $x$ and $N$, where $x$ represents 
the fraction of ``cooperators'' producing a public good and enhances the population growth rate through the
benefit $0<b\sim{\cal O}(1)$. Both strains compete for limited resources which constrains the population size 
as encoded by the death rate $T_\alpha^-=N/K$. We consider that in the presence of
 environmental randomness, $K$ fluctuates stochastically.
The population thus follows a multivariate birth-death process~\cite{Gardiner,KEM2} in which, at each time increment, 
an individual at random reproduces (with per-capita rate $g(x)T_\alpha^+$), or dies (with per-capita rate $T_\alpha^-$), 
or the carrying capacity changes state (with rate $\nu$).
The ensuing master equation fully describes the stochastic population dynamics, whose main features are the distribution of 
$N$ and the probability that $S$ or $F$ fixates by taking over the population, but is difficult to solve~\cite{KEM2}.
Upon ignoring any form of noise, the population size $N$ and composition $x$ evolve deterministically
according to~\cite{Melbinger2010,Cremer2011,SM}
\begin{eqnarray}
\label{eq:N-deterministic}
\dot{N}&=& N\left(g(x)-\frac{N}{K}\right)\,,\\
\label{eq:x-deterministic}
\dot{x}&=&-sg(x)\frac{x(1-x)}{1-sx}\,,
\end{eqnarray}
where the dot signifies the time derivative.
Here, we study  the population dynamics subject to a randomly switching 
carrying capacity (environmental noise) and to stochastic birth and death events 
(internal noise). We therefore have to account for these sources of noise.

To model environmental randomness, we let the carrying capacity $K(t)$ switch stochastically between a state of abundant resources ($K=K_+$) and one of scarcity ($K=K_-<K_+$).
Figure~\ref{fig:Cartoon} illustrates this stochastic environment and its impact on the population.
We consider that environmental switching occurs continuously at rate $\nu$, according to a 
dichotomous Markov noise $\xi(t)\in\{-1,+1\}$ with zero-mean, $\langle \xi(t)\rangle=0$ ($\langle \cdot\rangle$ denotes the 
ensemble average), and autocorrelations  $\langle \xi(t) \xi(t')\rangle={\rm exp}(-2\nu|t-t'|)$, where $1/(2\nu)$ is the
{\it finite correlation time}~\cite{HL06,Bena06}. 
Hence, the carrying capacity obeys
\begin{eqnarray}
 \label{eq:K(t)}
 K(t)=\frac{1}{2}\left[(K_+ + K_-)+\xi(t) (K_+ - K_-)\right]\,,
\end{eqnarray}
with average $\langle K\rangle=(K_+ + K_-)/2$.
If this is the sole source of noise (no internal noise), the evolution obeys a piecewise deterministic 
Markov process (PDMP)~\cite{PDMP1,PDMP2,Hufton16,Hidalgo17}, defined by (\ref{eq:x-deterministic}) 
and
\begin{eqnarray}
	\dot{N}&=& N\left\{g(x)-\frac{N}{{\cal K}}+
	\xi\frac{N(K_+-{\cal K})}{{\cal K}K_+}
	\right\}\,,\label{eq:N-stochastic}
\end{eqnarray}
where ${\cal K}=2K_+K_-/(K_+ + K_-)$ is the harmonic mean of $K_+$ and $K_-$. Equation (\ref{eq:N-stochastic}) is 
obtained  from  Eqs.~(\ref{eq:N-deterministic}) and (\ref{eq:K(t)}) as shown
in the Supplemental Material~\cite{SM}.
Hence, environmental randomness alone yields a multiplicative noise $\propto \xi (K_+ -K_-) N^2$ in (\ref{eq:N-stochastic}).
Demographic fluctuations being ignored, $x$ obeys Eq.~(\ref{eq:x-deterministic}), which is decoupled 
 from $N$, and  evolves on a  timescale $\sim 1/s$, see supporting videos \cite{movies} and Supplemental Material \cite{SM}.

Internal noise arises in finite populations when birth and death events occur randomly,
and is responsible for fixation.
If  demographic fluctuations are the only source of noise  (say $K$ is constant), the fixation 
probability $\phi$ of the strain $S$ can be computed from a fitness-dependent Moran 
process~\cite{Moran,Kimura,Ewens,Antal06} with the same strain-specific fitnesses as in our model, and constant size $N=K$~
\cite{Otto}.
Given an initial fraction $x_0$ of $S$ individuals, this probability in a population of constant size $N$ is 
$\phi(x_0)|_{N}= (e^{-N s(1-x_0)}-e^{-N s})/(1-e^{-N s})$~\cite{Blythe07,Cremer09}.
Hence, the fixation probability of the slow strain is  exponentially small in large size populations.
Since the fixation probability clearly depends on $x_0$, for notational simplicity
 we henceforth  write $\phi\equiv \phi(x_0)$ and $\phi|_{N}\equiv \phi(x_0)|_{N}$.

Below, we investigate the \textit{joint} effect of environmental and internal noise on the population dynamics.
In particular, since extreme environmental changes 
can occur more or less rapidly in microbial communities~\cite{Wahl02,Patwas09,Wienand15,Brockhurst07a,Brockhurst07b}, 
we study the influence of the switching rate $\nu$ on the species fixation probability and the distribution of $N$.

\begin{figure}[htb]
\includegraphics{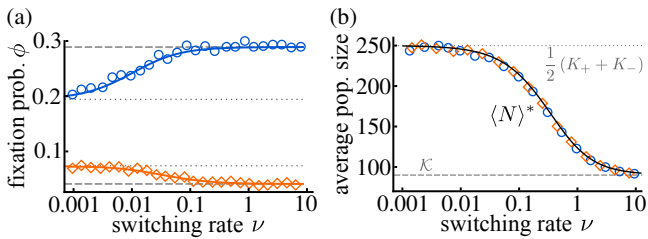}
\caption{{\it (Color online)}. 
(a) $\phi$ vs. $\nu$ for $(K_{+},K_-,x_0)=(450,50,1/2)$, with $s=0.02$ ($\circ$, blue/black) and $s=0.07$ 
($\diamond$, orange/gray). 
Symbols are from simulations ($10^4$ runs). Solid lines are from
 (\ref{phi-nu}); dashed/dotted lines show $\phi$ when 
$\nu/s \to \infty$ (dashed) and $0$, see text.
(b) $\langle N\rangle^*$ vs. $\nu$. Symbols are  from simulations ($10^4$ runs) with 
$s=0.02$ ($\circ$) and
$s=0.07$ ($\diamond$); they collapse on the curve (solid line) obtained 
by averaging  $N$  over (\ref{pstar}), see text.
}
\label{fig:Fig2}
\end{figure}

{\it (i) The pure resource competition scenario.}
When  $g=1$, both species simply compete for limited resources.
By the competitive exclusion principle~\cite{Hardin60}, $F$ always prevails in the deterministic limit.
In this case, the rate equations (\ref{eq:N-deterministic}),(\ref{eq:x-deterministic}) are decoupled.
However, demographic fluctuations, which drive to fixation, scale with the population size: the {\it stochastic dynamics} of $x$ is thus coupled to that of $N$, see Fig.~\ref{fig:Cartoon}.
While $x$ relaxes on a slow timescale $t\sim 1/s$, $N$ 
reaches a quasi-stationary state in a time  $t={\cal O}(1)$, see supporting videos \cite{movies} and Supplemental Material \cite{SM}. 
Eq. (\ref{eq:N-stochastic}) is associated with a PDMP whose marginal (unconditioned of $\xi=\pm 1$) stationary 
probability density function (PDF) is~\cite{HL06,SM}
\begin{eqnarray}
\label{pstar}
\hspace{-7mm}
p_{\nu}^*(N)= \frac{{\cal Z}_{\nu}}{N^2}~\left[\frac{(K_{+} -N)(N-K_{-})}{N^2}\right]^{\nu-1},
\end{eqnarray}
where  ${\cal Z}_{\nu}$ is the normalization constant and
the PDF has support $[K_{-},K_{+}]$.
Although this PDF only accounts for environmental noise, it captures the main features of the quasi-stationary distribution 
of the population size ($N$-QSD) of the full model when $K_-\gg 1$~\cite{meta}.
Since $x$ and $N$ evolve on different timescales, the PDF (\ref{pstar}) can
be combined with $\phi|_{N}$ to determine the fixation probability. For this, we
rescale the switching rate, $\nu\to\nu/s$, to map  environmental changes onto the internal dynamics' relaxation timescale,
where $\nu/s$ is the average number of switches occurring while $x$ relaxes. 
Indeed, when $\nu\gg s$ (fast switching), many switches occur
prior to fixation and the environmental noise self-averages, whereas when $\nu\ll s$ (slow switching)
the population is likely to solely experience the carrying capacity  $K_{+}$ or $K_{-}$  before one species fixates.
The fitness-dependent Moran process gives the  fixation probability
in those limits. When $\nu\to \infty$, there is self-averaging with
$\xi \to \langle \xi\rangle =0$ in
 (\ref{eq:N-stochastic}) that becomes the logistic equation (\ref{eq:N-deterministic}) with
 $K={\cal K}$, yielding $\phi=\phi|_{\cal K}$. When $\nu \to 0$, $K$ is equally likely to remain at 
 $K_{+}$ or $K_{-}$ until fixation occurs, yielding $\phi=(\phi|_{K_+}+\phi|_{K_-})/2$.
Based on these physical considerations, fully detailed in Supplemental Material~\cite{SM}, 
we propose to assume the following expression for the $S$ fixation probability when $0<s\ll 1$ and $K_-\gg 1$: 
\begin{eqnarray}
\label{phi-nu}
\phi&\simeq& \int_{K_-}^{K_+} \left(\frac{e^{-N s(1-x_0)}-e^{-N s}}{1-e^{-N s}}\right)~p_{\nu/s}^*(N)~dN.
\end{eqnarray}
By averaging the effect of internal noise, given by $\phi|_N$, over the external-noise-induced PDF $p_{\nu/s}^*$, 
Eq.~(\ref{phi-nu}) accounts for the fact that $N$ evolves much faster than $x$ relaxes. 
The expression (\ref{phi-nu}) reproduces the expected results in 
the two limiting regimes $\nu\gg s$ and $0<\nu\ll s$. Moreover, (\ref{phi-nu}) accurately predicts the 
stochastic simulation results over a broad range of $\nu$ values,
capturing the nontrivial $\nu$-dependence of $\phi$, see Fig.~\ref{fig:Fig2}(a).
We find that $\phi$ can increase or decrease with $\nu$~\cite{SM} and,  importantly, 
environmental noise can significantly enhance the $S$ fixation probability in {\it all} regimes: 
$\phi$ is always  greater than  $\phi|_{\langle K\rangle}$ obtained in a  
non-random environment with $N=\langle K\rangle$~\cite{SM}.
\begin{figure}[htb]
\includegraphics{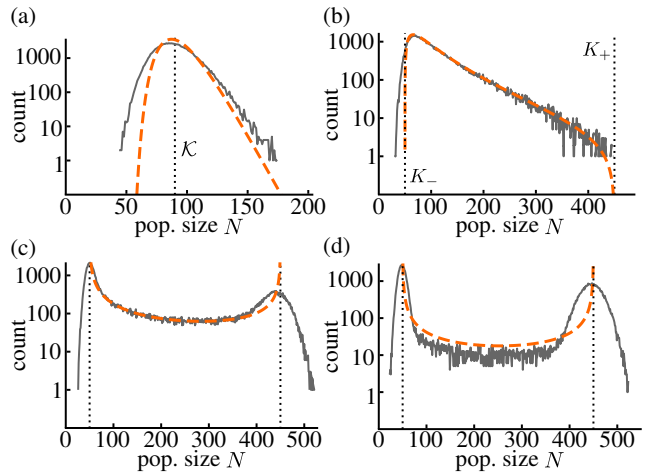}
\caption{{\it (Color online)}. Histograms of population size
($N$-QSD) and  from $p_{\nu}^*$, for $\nu=20$ (a), $\nu=2$ (b),
$\nu=0.2$ (c), and $\nu=0.02$ (d).
Solid lines result from simulations ($10^5$ samples, after $t\gtrsim 1/s$).
Dashed lines are the corresponding histograms from Eq.~(\ref{pstar}).
Dotted lines show $N={\cal K}$ in (a), and $N=K_{\pm}$ in (b)-(d).
Parameters are  $(K_+, K_-,s,x_0)=(450, 50,0.02,0.5)$.}
\label{fig:Fig3}
\end{figure}

We have verified that the mean fixation time scales as ${\cal O}(1/s)$~\cite{SM}. 
Hence, after a time $t \gtrsim 1/s$, either species likely fixated and,
while the population then only consists of $S$ or $F$, its size keeps fluctuating, see supporting  videos \cite{movies} and Supplemental Material \cite{SM}.
Since demographic fluctuations have a marginal influence on the $N$-QSD when $K_-\gg 1$, 
the PDF $p_{\nu}^*$  captures its main long-time features, see Fig.~\ref{fig:Fig3}. 
For example, 
the long-time average population size $\langle N \rangle^*$ is well described by 
the average over Eq.~(\ref{pstar}): 
$\langle N \rangle^* \simeq \int_{K_-}^{K_+} N p_{\nu}^*(N) dN$, which  
is independent of $s$ and $x_0$, see Fig.~\ref{fig:Fig2}~(b).
The histograms of Fig.~\ref{fig:Fig3}  show that the environmental noise causes a  {\it noise-induced}
transition of the $N$-QSD at about $\nu =1$~\cite{HL06}. The transition, predicted by $p_{\nu}^*$, separates regimes
in which  environmental change is faster or slower  than the population's growth rate.
For $\nu>1$, fast  switching results in a unimodal $N$-QSD, see Fig.~\ref{fig:Fig3}~(a,b), whereas for 
$\nu<1$, the environment changes slowly and the $N$-QSD is bimodal, see Fig.~\ref{fig:Fig3}~(c,d) and \cite{SM}.
The fast decay and slower growth of $N$, characteristic of a logistic dynamics,
lead the population size to dwell longer about $K_-$ than about $K_+$. As captured by $p_{\nu}^*$, this results in right-tailed distributions in Fig.~\ref{fig:Fig3}.
 Since (\ref{pstar}) only accounts for external noise, it cannot reproduce some features caused by 
 demographic fluctuations, such as the $N$-QSD  not being strictly confined within the support of $p_{\nu}^*$~\cite{SM,KEM2}.
However, as Fig.~\ref{fig:Fig2} shows, these deviations only marginally affect $\langle N \rangle^*$ and $\phi$.

\begin{figure}
\includegraphics{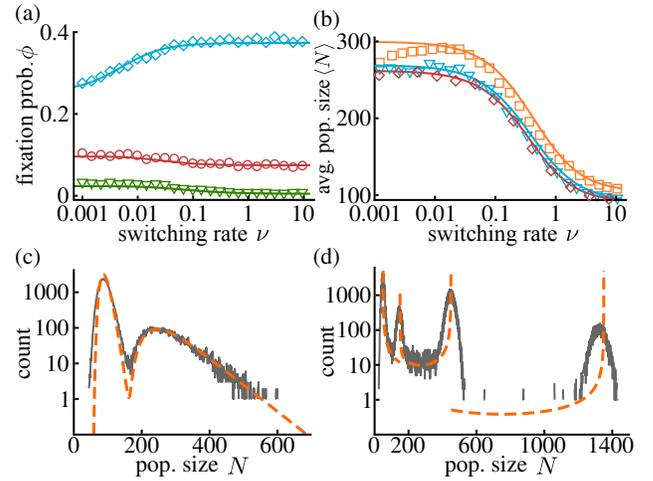}
\caption{
{\it (Color online)}. (a)  $\phi$ vs $\nu$  for 
$(s,b)=(0.01,0.2)(\diamond$, blue/gray),~$(0.05,0.2)(\circ,$ red/black),~$(0.05,2)(\nabla,$ green/dark gray).
Solid lines are 
from (\ref{phi-nu_q}). In all panels $(K_{+},K_-,x_0)=(450,50,0.5)$.  
(b) $\langle N \rangle^*$ vs. $\nu$  for $(s,b)=(0.025,2)$ ($\square$, orange/gray),~$(0.05,2)$ ($\triangledown$, 
blue/dark gray),~$(0.025,8)$ ($\diamond$, red/black). Solid lines are from (\ref{eq:Nstarq}). 
(c,d) Size distributions for $ \nu=20$ (c) and $ \nu=0.02$ (d), with $b=2$ and $s=0.02$.
Solid and dashed lines are respectively histograms
from simulations ($10^5 $ replicas, after 99$\%$ fixation~\cite{SM})
and obtained from $p_{\nu,b}^*$ and $p_{\nu}^*$ weighted by $\phi_q$, see text.
}
\label{fig:Fig4}
\end{figure}

{\it (ii) The public good scenario.}
The above approach can  be generalized to cover cases where internal and ecological dynamics are explicitly coupled.
As an application, we consider a public good scenario in which $S$ is a ``cooperative'' strain benefiting 
the population by enhancing the global fitness $g(x)=1+bx$ ($b>0$) and 
the carrying capacities, see below.
The dynamics of $x$ and $N$ are now coupled, breaking the timescale separation: 
$N$ becomes a fast variable,
enslaved to the slowly-varying $x$, see videos 6 and 7 in Ref.~\cite{movies}  and Supplemental Material \cite{SM}.
After fixation,  $x\in\{0,1\}$
and the $N$-QSD  can be obtained as for $b=0$.
When $F$ fixates ($x=0$), the $N$ distribution is described by $p^*_{\nu}$ (\ref{pstar}). 
If $S$ fixates ($x=1$), the population size distribution
is captured by $p^*_{\nu,b}$, obtained by substituting $K_\pm\to(1+b)K_\pm$ and $\nu\to\nu/(1+b)$ in Eq.~(\ref{pstar}). 
Hence,  $p^*_{\nu}$ and $p^*_{\nu,b}$ are  the PDFs conditioned to fixation of $F$ and $S$ (but unconditioned of $\xi$), 
respectively.
To address the dynamics before fixation, we approximately account for the correlations between $N$ and $x$ by introducing
an effective (constant) parameter $0\leq q\leq b$. We then set $g(x)=1+q$  in (\ref{eq:N-stochastic}), resulting in a PDMP, 
decoupled from $x$, for the size of an effective population
whose  marginal PDF, $p^*_{\nu,q}$ (see Eq.~(S2) in \cite{SM}), interpolates between $p^*_{\nu}$ and $p^*_{\nu,b}$.
As for $b=0$, when $0<s\ll 1$ and $K_-\gg 1$, the $S$
fixation probability in this effective population is~\cite{SM}
\begin{eqnarray}
\label{phi-nu_q}
\hspace{-5mm}
\phi_{q}&=& \int_{(1+q)K_{-}}^{(1+q)K_{+}}\!\! \left(\frac{e^{-N s(1-x_0)}-e^{-N s}}{1-e^{-N s}}\right) 
p_{\nu/s,q}^*(N)~dN.
\end{eqnarray}
To determine the effective value of $q$ for  given  $(K_\pm,s,b)$, we consider 
the limit $\nu\gg 1$, where the environmental noise self-averages,
and match the prediction of (\ref{phi-nu_q}) with the  fixation probability obtained in simulations~\cite{SM}.
As Fig.~\ref{fig:Fig4}(a) shows, 
with suitable $q$, (\ref{phi-nu_q}) reproduces the simulation results, $\phi_q \simeq \phi$, for a broad range of $\nu$ and 
different $b$ values.

After $t\gtrsim 1/s$, fixation has typically occurred and the population size 
distributions (when $K_-\gg 1$) are well 
described by  $p_{\nu,b}^*$ ($S$ fixation) and $p_{\nu}^*$ ($F$ fixation).
With these conditional PDFs and  $\phi_q$, the long-time average population size reads
\begin{eqnarray}
 \label{eq:Nstarq}
 \hspace{-3mm}
 \langle N \rangle^*\simeq \phi_q\int_{(1+b)K_-}^{(1+b)K_+}\!\!\!\! N p_{\nu,b}^*(N) dN +
 \widetilde{\phi}_q\int_{K_-}^{K_+} N p_{\nu}^*(N) dN,
\end{eqnarray}
with  $\widetilde{\phi}_q=1-\phi_q$. 
Fig.~\ref{fig:Fig4}(b) shows that (\ref{eq:Nstarq}) agrees well with simulation results, but 
cannot capture the behavior at very low $\nu$ ($\phi_q$ being inferred at $\nu\gg1$).
The conditional $N$-QSD  and conditional PDFs $p_{\nu}^*$ and $p_{\nu,b}^*$ 
present unimodal and bimodal regimes. Specifically, after $S$ fixation, $N$'s growth rate is $1+b$ and the 
associated PDF $p_{\nu,b}^*$ undergoes 
the
noise-induced transition at $\nu=1+b$. Similarly, the $N$'s growth rate
when $F$ fixates is $1$, and  $p^*_\nu$ undergoes a transition at $\nu=1$.
Since the marginal size distribution is the sum of  the conditional distributions weighted by the fixation probability, 
it is characterized by several regimes and  transitions. These properties  are well captured by combining 
$p_{\nu,b}^*$ and $p_{\nu}^*$ weighted by $\phi_q$, as shown in Fig.~\ref{fig:Fig4}.
When $\nu>1+b$, the switching rate exceeds the population's growth rate, and
both conditional PDFs  are unimodal with different peaks, yielding a bimodal 
marginal distribution, see  Fig.~\ref{fig:Fig4}(c).
For $1<\nu<1+b$,  $p_{\nu,b}^*$ is bimodal and $p_{\nu}^*$ is unimodal, 
When $\nu$ is below the  population's growth rate ($\nu<1$), both conditional PDFs are bimodal.
As a result, the marginal size distribution has three peaks when $1<\nu<1+b$ and 
four peaks when $\nu<1$, see  Fig.~\ref{fig:Fig4}(d).
As for $b=0$, the influence of demographic fluctuations on the $N$-QSD is to cause slight deviations
 from the PDF predictions, particularly at low $\nu$~\cite{SM}.

Motivated by the evolution of microbial communities in volatile environments, we have analyzed the dynamics of 
a two-species population subject to a randomly switching carrying capacity (dichotomous noise).
A distinctive feature of our model is the coupling of the environmental and internal noise: 
demographic fluctuations  depend on the population size, which in turn changes with the 
varying carrying capacity (environmental noise).  By analytical and computational means, we have studied 
the coupled effect of  environmental and internal noise on the population's ecological 
and fixation properties. Our analytical approach is based on a timescale separation, arising under weak selection, between the
ecological and internal dynamics. We have also combined the properties of suitable  stochastic processes
governed solely by internal fluctuations on one hand, and only by environmental noise on the other hand.
In the case of pure resource competition (no explicit coupling between internal and ecological 
dynamics),  we have determined the population size distribution, characterized by various regimes, and found that the average size 
decreases with the  switching rate.
Assuming a suitable expression for the fixation probability and using stochastic simulations, we have investigated how environmental 
randomness affects the strains’ fixation properties and found that it can significantly enhance
the fixation probability of the disadvantaged strain.
As an application, we have considered a public good scenario in which internal and ecological dynamics
are explicitly coupled. We have thus devised an effective theory that has  allowed us to probe the effects of 
environmental switching and public good benefit on the fixation probability and population composition.
We have  characterized the population size distribution and
the noise-induced transitions between their unimodal (fast switching), 
bimodal and multimodal forms,
arising when the switching rate matches that of the population 
growth.
Our  findings show that coupled environmental and demographic noise can significantly influence  
the population dynamics by greatly affecting its fixation properties and therefore its composition.
This is particularly relevant to  microbial communities, which often feature connected internal 
and ecological evolution.

EF acknowledges funding by the Deutsche Forschungsgemeinschaft, Priority Programme 1617 ``Phenotypic heterogeneity 
and sociobiology of bacterial populations'', grant FR 850/11-1,2, and the German Excellence Initiative via the program 
``Nanosystems Initiative Munich'' (NIM). MM is grateful for the  support of the Alexander von Humboldt Foundation, 
Grant No. GBR/1119205 STP, and for the hospitality of the University of Munich.

\renewcommand{\thefootnote}{\fnsymbol{footnote}}
\renewcommand{\thefigure}{S\arabic{figure}}
\renewcommand{\theequation}{S\arabic{equation}}
\setcounter{figure}{0}
\setcounter{equation}{0}

\newpage

\begin{widetext}

\begin{center}
{{\itshape\Large Supplemental Material for}\\~\\\Large Evolution of a Fluctuating Population in Randomly Switching Environment}
\end{center}

In this Supplemental Material, we provide some  technical details and supplementary information in support of the results discussed 
in the main text, as well as additional ones concerning the population's mean fixation time and its long-time distribution.
We also comment the content of electronically available Videos~\cite{movies} that illustrate the population dynamics in the 
pure resource competition and public good scenarios. In what follows, unless otherwise stated, the notation is the same as in the main text and the equations and
figures refer to those therein. (As in the main text, unless stated otherwise, below we tacitly assume $x_0=1/2$.)

\vspace{0.5cm}

\section{1. Derivation of Equations (1,2) and (4)}
As explained in the main text, the population dynamics is governed by multivariate birth-death process in which at each time increment  an  individual of species $\alpha\in \{S,F\}$
is picked for reproduction, $N_{\alpha}\to N_{\alpha}+ 1$, with transition rate ${\cal T}_{\alpha}^{+}=T_{\alpha}^{+}N_{\alpha}=g(x)
f_{\alpha}N_{\alpha}/\bar{f}$ or death, $N_{\alpha}\to N_{\alpha}- 1$, with transition rate ${\cal T}_{\alpha}^{-}=T_{\alpha}^{-}N_{\alpha}=(N/K)N_{\alpha}$, 
or  the carrying capacity is switched, $K_+ \leftrightarrow K_-$,  with rate $\nu$.
 When internal noise is neglected, $N$ and $x$ evolve according to the mean-field rate equations
\begin{eqnarray*}
	\dot{N}&=& \sum_{\alpha=S,F}({\cal T}_{\alpha}^{+}-{\cal T}_{\alpha}^{-}) =N\left(g(x)-\frac{N}{K}
	\right),\label{RE1}\nonumber\\
	\\
	\dot{x}&=&\frac{{\cal T}_{S}^{+}-{\cal T}_{S}^{-}}{N}-x\frac{\dot{N}}{N}=-sg(x)~\frac{x(1-x)}{1-sx}
	\label{RE2},\,
\end{eqnarray*}
where we have used $f_F=1$, $f_S=1-s$ and $\bar{f}=1-sx$. These equations coincide with (1) and (2) and, when 
 the carrying capacity $K$ is constant, they provide a suitable description of the  ecological and evolutionary (composition).
The deterministic description of the population dynamics in terms of (1) and (2) is valid only
 in the absence of internal and external noise. 
\\ When the carrying capacity  randomly switches according to 
 $K(t)=\frac{1}{2}\left[(K_+ + K_-)+\xi(t) (K_+ - K_-)\right]$, where $\xi\in \{-1,+1\}$ is the dichotomous noise 
 defined in the main text, the  equation for $N$ becomes the following stochastic differential equation obtained
by substituting $K(t)$ into (1) and using $\xi^2=1$:
\begin{eqnarray*}
	\dot{N}&=& N\left(g(x)-\frac{2N}{K_+ + K_- +\xi(t) (K_+ - K_-)}\right)=
	N\left(g(x)-\frac{N}{2K_+ K_-}[K_+ + K_- -\xi(t) (K_+ - K_-)]\right)\nonumber\\
	&=& N\left(g(x)-\frac{N}{{\cal K}}+\xi N\left\{\frac{K_+ + K_-}{2K_+ K_-}-\frac{2K_-}{2K_+ K_-}\right\}\right)=
	N\left(g(x)-\frac{N}{{\cal K}}+
	\xi\frac{N(K_+-{\cal K})}{{\cal K}K_+}\right),
\end{eqnarray*}
where  ${\cal K}=2K_+ K_-/(K_+ + K_-)$. This stochastic differential equation coincides with (4) 
and, together with (2), defines a piecewise deterministic Markov process (PDMP)~\cite{PDMP1,PDMP2,HL06} describing 
the population dynamics when the sole form of randomness is the random switching of the carrying
capacity (internal noise is  neglected).

\section{2. Fixation probability under random switching: Arguments underpinning formula (6) and (7) and their properties}
We have studied the fixation probability $\phi$ that, starting with a fraction $x_0$ of individuals of
the slow type $S$, the entire population eventually consists of $N(t)$ individuals of species $S$. The fixation of species 
$F$ occurs with the complementary probability $\widetilde{\phi}=1-\phi$. 
We have investigated the joint effect of external (dichotomous) and internal (demographic) noise on these fixation 
probabilities with help of Eqs.~(6) and (7)  when $K_- \gg 1, 0<s\ll 1$ and $\nu>0$, and by comparing
the predictions of these formula with the results of stochastic simulations carried out using the Gillespie algorithm~\cite{Gillespie} which 
{\it exactly} simulates the master equation.

\subsection{2.1 Physical arguments underpinning formula (6) and (7) and their corroboration}
Formula (6) and (7) are assumed forms for the fixation probability $\phi$ of the slow species $S$ when $0<s\ll 1$ and $K_-\gg 1$.
These expressions are based on a series of physical considerations that are fully corroborated
by stochastic simulations of the underlying individual-based population dynamics.
At its core, the rationale behind (6) is rooted in the timescale separation between $N$ and $x$ and on scaling arguments.
For the sake of concreteness, here we first focus on the case of pure resource competition ($b=0$) 
and present the physical arguments underpinning Eq.~(6):
\begin{enumerate}
 \item[-] The condition  $0<s\ll 1$ ensures that there is a timescale separation between the evolutionary and ecological dynamics.
In fact, as shown in the Videos 1-3~\cite{movies},   $x$ evolves on a much slower timescale than $N$ when $0<s\ll 1$: $x$ relaxes
in a time of order $1/s$ while $N$ is at quasi-stationarity after a time of order 1. The condition  $K_- \gg 1$ 
ensures that the evolution of the population size is chiefly driven by random switching and is 
well described by the PDMP (4) that neglects the effects of demographic noise that are marginal when $K_- \gg 1$ 
(see also Sec. 3 below).
\item[-] Due to the timescale separation,  when fixation occurs, typically after a time of order $1/s$ (see Fig.~S3), $N$ 
can be considered to be in the stationary state of the PDMP (4) whose probability density function 
(PDF) has support $[K_-, K_+]$. 
\item[-] The evolution of $x$ is much slower than the dynamics of $N$. The population size is therefore able to span much of its quasi-stationary distribution before fixation.
This suggests to (approximately) compute $\phi$ by averaging  $\phi|_{N}$, which is the $S$ fixation probability in a fitness-dependent Moran model 
of constant population size $N$ (see main text), over the stationary PDF of the underlying PDMP that captures the main features of long-time dynamics of $N$.
\item[-] Since $x$ evolves on a timescale $1/s$ times slower than $N$, when  $1/\nu$ (mean time between two switches) is much 
shorter than  $x$'s relaxation time,
the population composition changes by $1/N$ while $N$ has already typically experienced many switches. Hence, 
when  $\nu\gg s$, the external noise self averages on the 
timescale of the relaxation of $x$ even if $N$ experiences large excursions 
 ({\it e.g.}, from $N\approx K_{\pm}$ to  $N\approx K_{\mp}$ as in the case of Fig.~3(c)): Hence,
$x$ changes by $1/N$ while the  population size $N$ appears to fluctuate about  a characteristic value.
It is therefore necessary to rescale the switching 
rate $\nu \to \nu/s$ in  averaging $\phi|_{N}$ over the stationary PDF of Eq.~(4) in order to compute the fixation probability $\phi$. The rescaling  $\nu \to \nu/s$ reflects
the fact that  $K(t)$ experiences on average $\nu/s$ switches prior to fixation (while $x$ relaxes).
In other words, this means that in this context the extent to which the environmental noise self-averages relative to the typical 
relaxation time of $x$ determines whether the environment changes
``fast'' or ``slowly''.
\item[-] With this rescaling, we obtain
 Eq.~(6): $\phi\simeq \int_{K_-}^{K_+} \phi|_{N}~p_{\nu/s}^*(N) dN$, where 
 the integral over $N$ spans $[K_-, K_+]$ which is the support of $p_{\nu/s}^*(N)$
 given by Eq.~(5). In  Eq.~(6), $\phi|_{N}$ accounts for internal noise
 in a population of size $N$ while $p_{\nu/s}^*(N)$ 
 captures the effect of the environmental noise on the (quasi-)stationarity distribution of $N$
 in terms of the PDMP (4).
\item[-] In the fast and slow switching regimes, the fixation probability $\phi$ can be computed directly from the properties of 
the fitness-dependent Moran model. In fact, when  $\nu \to \infty$ (very fast switching),
the dichotomous noise  self-averages ($\xi\to \langle \xi \rangle=0$ in Eq.~(4)) and the population readily attains the 
the effective size $N\simeq {\cal K}\gg 1$. The  internal evolution thus
mirrors that of a population of constant size ${\cal K}$ obeying  a fitness-dependent
Moran process~\cite{Moran,Ewens,Cremer09}. In this case, if the initial fraction of $S$ individuals is 
$ x_0 $, we have  $\phi  \xrightarrow{\nu \to \infty} \phi^{(\infty)}=\phi|_{\cal K}= 
(e^{-{\cal K}s(1-x_0)}-e^{-{\cal K}s})/(1-e^{-{\cal K}s})$~\cite{Ewens}, see main text.
Similarly, when $\nu \to 0$ (very slow switching), the population is equally likely to be 
locked in either of the environmental state $\xi=-1$ (where $N=K_-$)
or $\xi=+1$ (where $N=K_+$) and from the properties of the fitness-dependent Moran model
in this case the fixation probability is
$\phi  \xrightarrow{\nu \to 0}  \phi^{(0)}=(\phi|_{K_-}+\phi|_{K_+})/2$.
\item[-] The stationary PDF $p_{\nu/s}^*(N)$ in Eq.~(6) accounts for the fact that  when $\nu>s$
there are typically many switches prior to fixation, and environmental noise essentially self-averages when 
$\nu\gg s$ and a large number of switches occur. In fact,  Eq.~(6) correctly reproduces
the fixation probability under fast and slow switching: it predicts $\phi\simeq \phi^{(\infty)}$  when $\nu/s\gg 1$
and $\phi\simeq \phi^{(0)}$  when $\nu/s\ll 1$, see Figs.~2(a) and 
S1.
\item[-] The stationary PDF  $p_{\nu/s}^*(N)$ is unimodal with a peak at
$N\approx {\cal K}$ when $\nu>s$,  and is bimodal  with peaks about $N=K_{\pm}$ when $\nu<s$, see Fig. 3 and Videos 4 and 5. This suggests
that in the regime of intermediate switching rate,  shown as shaded areas in Fig.~\ref{fig:phi-supp}, the fixation probability interpolates
between $\phi^{(0)}$ and $\phi^{(\infty)}$, and we expect that$\phi\approx \phi^{(\infty)}$ over a broad range 
of values of $\nu$ since $s\ll 1$ and $\nu/s\gg 1$ is always satisfied when $\nu$ is of order 1.

\begin{figure}[h]
\begin{center}
\includegraphics{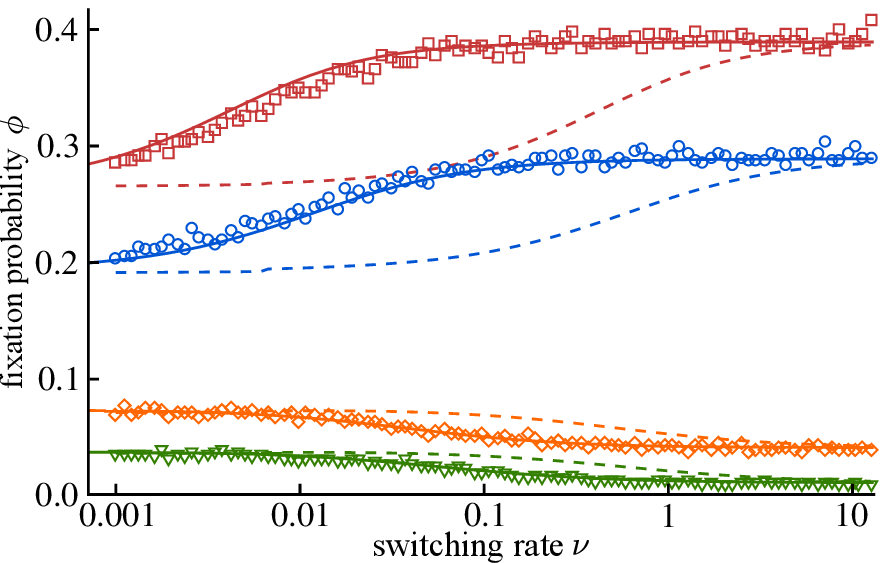}
\end{center}
\caption{{\it (Color online)}. 
Fixation probability $\phi$ as function of $\nu$ for different values of $s$ in the pure competition case ($b=0$).
Here, $(K_+,K_-)=(450,50)$.
Symbols denote the results of stochastic simulations for different values of $s$:
$s=0.01$ ($\square$, red/ dark gray), $s=0.02$ ($\circ$, blue/black), 
$s=0.07$ ($\diamond$, orange/gray) and $s=0.1$ ($\triangledown$, green/light gray), from top to bottom. Solid lines denote the corresponding predictions of Eq. (6) and dashed lines represent the predictions 
of $\int_{K_-}^{K_+} \phi|_{N}~p_{\nu}^*(N) dN$, obtained by averaging $\phi|_{N}$
over (5) {\it without} rescaling the switching rate $\nu$, see text.
The rescaling of the switching rate into $\nu/s$ in Eq.~(6) reveals that 
$\phi$ is a scaling function of $\nu/s$. In fact, without rescaling the switching rate,
the predictions (dashed lines) agree with stochastic simulations only in the regimes of very 
large $\nu$ (fast switching) or very low $\nu$ (slow switching); whereas the predictions of Eq.~(6)
agree with simulations over four orders of magnitude.
Hence, Eq.~(6) with the average over the PDF (5) with rescaled switching rate $\nu \to \nu/s$ provides  accurate predictions in the 
intermediate switching regime that separates  the slow ($\nu/s  \ll 1$) and fast ($\nu/s  \gg 1$) switching regimes, see text.
}
\label{fig:phi-supp}
\end{figure}

\end{enumerate}
At this point, it is worth emphasizing that the assumed form Eq.(6) has been proposed 
{\it without making any use of fitting parameters and does not rely on any  input from stochastic simulations}, but only
on the basis of the above physical considerations. 
Stochastic simulations have  been used to validate the form of (6) by corroborating its predictions. In fact, a pragmatic and efficient way to  assess the validity/accuracy of (6)
is to systematically compare its predictions with results of extensive stochastic simulations of 
 system's dynamics based on the Gillespie algorithm (typically sampling over $10^4$ to $10^5$ realizations). The form of Eq.~(6) and the above considerations 
are thus supported by the following evidence:
\begin{enumerate}
 \item[-] As shown in the supporting Videos 1-3~\cite{movies}, stochastic simulations fully confirm
 that $x$ always evolves much slower than $N$ when $s\ll 1$, and that a timescale separation occurs when $b=0$. 
 Figs 2-4, as well as the supporting Videos 4-5 also confirm that about the time of fixation 
(and after fixation has occurred), the  evolution of $N$ is well described by the underlying PDMP
when $K_- \gg 1$. In fact, except for the population collapse arising after an enormous, unobservable time, 
demographic noise has only a marginal effect on the (quasi-)stationary distribution of $N$.
 \item[-] Stochastic simulations mirroring the predictions of the system's master equation
 fully confirm that  Eq. (6) correctly predicts the expected behavior at fast and slow switching rate, with 
 $\phi  \simeq  \phi^{(\infty)}$ when $\nu/s\gg 1$
 and  $\phi  \simeq \phi^{(0)}$ when $\nu/s\ll 1$.  Furthermore, stochastic simulations
  show that the predictions of Eq.~(6) correctly reproduces the nontrivial $\nu$-dependence
 of $\phi$, see Figs. 2(a) and \ref{fig:phi-supp}, and 
  agree remarkably well with simulation results also in the regime of intermediate
 switching rate. 
 
The remarkable agreement between the predictions of Eq.(6) and 
stochastic simulations results has been confirmed for different values of $s$ (namely $s=0.01,0.02,0.07,0.1$),
and in all cases we have found an agreement within a few percent. More specifically, 
by a systematic comparison with simulations, we have estimated the mean square displacement of the predictions of Eq.~(6) from the simulation results to be 
within $1.5\%$ to $9\%$ for the results of Fig.~\ref{fig:phi-supp}, with an accuracy that increases when $s$ is lowered: In the tested datasets, the mean error
ranges from about  $1.5\%$ when $s=0.01$ to about $9\%$ when $s=0.1$ and $(K_-,K_+)=(50,450)$~\cite{KEM2}. The fact that the accuracy of 
(6) improves when $s$ is lowered stems from the fact that Eq. (6) is built on assuming a timescale separation between $N$ and $x$, which
is the more pronounced the lower $s$.
\item[-]  Gillespie stochastic simulations confirm that rescaling $\nu\to \nu/s$ is necessary to correctly predict the 
fixation probability on a broad spectrum of $\nu/s$ values.
This is illustrated in Fig.~\ref{fig:phi-supp}.
When we compare the predictions of Eq.~(6), obtained by averaging over the PDF (5)
with the rescaled switching rate $\nu/s$, against stochastic simulations for different values of $s$ we find an 
excellent agreement over the entire range of $\nu$ values (spanning four orders of magnitude, from $\nu\sim 10^{-3}$ 
to $\nu\sim 10$) 
On the other hand, the predictions of $\int_{K_-}^{K_+} \phi|_{N}~p_{\nu}^*(N) dN$,
in which the switching rate has not been rescaled, are shown to be at odds with 
the results of stochastic simulations when $0.01\lesssim \nu \lesssim 1$
which includes intermediate switching regime
(the agreement is restricted to a limited range of 
very large/small values of $\nu/s$ corresponding to the very fast/slow switching regimes).
\end{enumerate}

In summary, the results reported in Figs. 2(a) and S1 show that the predictions of Eq.~(6) are 
in excellent agreement with the outcome of the system's Gillespie simulations (mirroring 
the dynamics described by the master equation)
over a broad range of values of $\nu/s$ values. This 
confirms that Eq.~(6) is indeed a good assumed expression (or, by a slight abuse of language, a suitable ``Ansatz'') for the actual fixation probability. The difference between the predictions of (6) and the corresponding 
simulation results can be  estimated numerically, but for the purpose of our discussion here, it suffices to notice 
that an agreement within a few percent is found over the broad range of $\nu/s$ values tested. Further technical details 
about the accuracy of (6) will be investigated elsewhere~\cite{KEM2}.

\vspace{0.15cm}

The physical considerations leading to Eq.~(6) when $b=0$ also lead to Eq.~(7) in the public good scenario with $b>0$. 
However, since Eqs.~(2) and (4) for $N$ and $x$ are coupled in this case, we  use a constant effective parameter $q\geq 0$ in our analysis.
As explained in the main text (see also below), this parameter is determined by matching simulation results. In fact,
when $b>0$, the effective parameter $q$ is introduced by considering the auxiliary stochastic differential equation obtained by substituting $g=1+q$
in Eq.~(4), see Sec.~1, which yields
\begin{equation}
\label{SDEaux}
 \frac{\dot{N}}{N}=1+q-\frac{N}{K}=1+q-\frac{N}{{\cal K}}+\xi \frac{N(K_+-{\cal K})}{{\cal K}K_+}.
\end{equation}
 This equation is
decoupled from the rate equation (2) for $x$ and corresponds to a PDMP~\cite{PDMP1,HL06}, describing how the size of an 
{\it effective} population evolves under the sole effect of the environmental noise. This PDMP is characterized by a probability 
$p^{\pm}_{\nu,q}(N,t)=p_{\nu,q}(N,\xi=\pm 1,t)$  to be in state $\{N, \xi\}$ 
at time $t$ for $q$ given, where 
\begin{equation*}
\frac{\partial}{\partial t} p_{\nu,q}^{\pm}(N,t)=
-\frac{\partial}{\partial N} \left[N\left(1+q-\frac{N}{{\cal K}}\right) p_{\nu,q}^{\pm}(N,t)\right]-
\nu[p_{\nu,q}^{\mp}(N,t)-p_{\nu,q}^{\pm}(N,t)].
\end{equation*}
By assuming that the probability current is zero at $N=(1+q)K_{\pm}$ (natural boundary conditions~\cite{HL06}) and $\nu>0$,
the  stationary marginal  probability density function  
$p_{\nu,q}^*(N)={\rm lim}_{t \to \infty} ({p}_{\nu,q}^{+} (N,t) + {p}_{\nu,q}^{-}(N,t))$ of (\ref{SDEaux}) is 
given by~\cite{PDMP1,HL06,Bena06} 
\begin{eqnarray}
\label{pstarq}
p_{\nu,q}^*(N)= \frac{{\cal Z}_{\nu,q}}{N^2}~\left[\frac{\left\{(1+q)K_{+}-N\right\}\left\{N-(1+q)K_{-}\right\}}{N^2}\right]^{\frac{\nu}{1+q}-1},
\end{eqnarray}
where  ${\cal Z}_{\nu,q}$ is the normalization constant,
$(1+q)K_{\pm}$ are the effective carrying capacities, whose harmonic mean is $(1+q){\cal K}$, 
and $[(1+q)K_{-},(1+q)K_{+}]$ is the support of $p_{\nu,q}^*$.
\\
To determine the parameter $0\leq q\leq b$, we consider the limit $\nu \to \infty$.
In such a regime, the environmental noise switches very rapidly and  self-averages, and Eq.~(\ref{SDEaux}) is
thus characterized by an effective population size $N= (1+q){\cal K}$. The corresponding fixation probability of species $S$ 
is thus $\phi|_{(1+q){\cal K}}$.
We then vary $q$ in order to match $\phi|_{(1+q){\cal K}}$ with the fixation probability obtained in our 
simulations for $\nu \gg 1$~\cite{footnote}.\\
In the realm of this effective theory, we can use this $q$ to determine $p_{\nu,q}^*(N)$ given by (\ref{pstarq}). 
Then, as we did to obtain Eq.(6), an expression $\phi_q$ fixation probability of $S$ is obtained by averaging
$\phi|_N$ over (\ref{pstarq}) with a rescaled switching rate $\nu \to \nu/s$. This yields Eq.~(7)
for $\phi_q$ whose expression  has been used  in Figs.~4(b)-4(d) in lieu of $\phi$, see also Sec.~4 below. It is worth noting
that in the realm of this effective theory, the parameter $q$ accounts for the correlations of the dynamics of $x$ and $N$.\\
By setting $q=b>0$ in (\ref{pstarq}), we can obtain the (marginal) PDF $p_{\nu,b}^*(N)$ conditioned to the fixation of 
species $S$ (but unconditioned of whether $\xi=\pm 1$) in the public good scenario. Similarly, by setting $q=0$ in (\ref{pstarq}), we obtain
 $p_{\nu,0}^*(N)=p_{\nu}^*(N)$ which coincides with (5) and is the marginal PDF conditioned to the fixation of $F$  (but unconditioned of $\xi=\pm 1$) 
in the public good scenario and the marginal PDF in the pure resource scenario. In the latter case, 
$p_{\nu}^*(N)$ is used to obtain the expression (6) for the fixation probability $\phi$.

\subsection{2.2 Properties of formula (6) and (7)}
It is worth noting that formula (6) and (7) explicitly reflect  the coupling between internal and external noise.
\vspace{2mm}
\\
As discussed above, Eq.~(6) provides an excellent approximation of the fixation probability of $S$ for all the values of $\nu>0$, 
when $K_- \gg 1$.
Moreover, it captures the fact that external and internal noise can jointly significantly enhance the 
fixation probability of the slow type with the respect to its counterpart in a population of 
constant size $\langle K\rangle=(K_+ +K_-)/2\gg 1$ subject to  non-random environment, where this probability is exponentially small 
($\phi|_{\langle K\rangle}\approx e^{-\langle K\rangle s/2}$ when $x_0=1/2$ and $\langle K\rangle s \gg 1$).
This is also true in the limit $\nu \to 0$ where the population is as likely to be subject to a
carrying capacity smaller or larger than $\langle K\rangle$, which generally greatly increases the fixation probability of $S$ 
with respect to the case where $N=\langle K\rangle$ even if there may be no switches prior to fixation.
For instance, in Fig.~2(a)  we find that  $\phi\approx 0.20 \-- 0.30$
when $s=0.02$ while  $\phi|_{\langle K\rangle=250}\approx 0.08$, and for $s=0.07$ we have obtained 
$\phi\approx 0.05 \-- 0.07$ while  $\phi|_{\langle K\rangle=250}\approx 0.002$.
\vspace{2mm}
\\
Fig.~4(a) shows that expression (7) of $\phi_{q}$ is very close to $\phi$ when $\nu/s\gg 1$ (high switching rate) and $K_- \gg 1$, but slightly 
deviates from it when $\nu/s\ll 1$. This stems from the fact that the effective theory underpinning (7) builds on 
the value of $q$ inferred at high switching rate.
\vspace{2mm}
\\
Remarkably, both (6) and (7) are able to capture the nontrivial dependence of $\phi$ on the switching rate $\nu$, see Figs. 2(a), \ref{fig:phi-supp} and 4(a): 
$\phi$  increases with $\nu$ when $\phi^{(\infty)}>\phi^{(0)}$ 
and decreases when $\phi^{(\infty)}<\phi^{(0)}$. The former situation arises under sufficiently low selection pressure, whereas the 
latter scenario occurs above a certain selection intensity. The intuitive explanation for this is that $\nu\approx 0$
corresponds to a high-volatility-high-reward setting, in which $S$ is equally likely to end up in an environment 
with relatively high demographic noise ($K=K_-$), where its fixation probability is high, or in one ($K=K_+$) with 
low noise and lower fixation probability.
When  $\nu\gg 1$, on the other hand, the species $S$ is in a low-volatility-low-reward setting: it faces an 
almost constant population size ($N\approx {\cal K}$). When the selection intensity $s$ is increased, it becomes increasingly less 
favorable for $S$ to be in the low-volatility-low-reward setting, and thus $\phi^{(\infty)}<\phi^{(0)}$ and thus $\phi$ decreases with $\nu$.
In the case of Fig. 2(a), we can explicitly determine the critical selection pressure $s_c$ below which 
$\phi^{(\infty)}>\phi^{(0)}$.
When $K_{+}\gg K_-\gg 1$, we have 
${\cal K}=2K_-(1+{\cal O}(K_-/K_+))$ and therefore $\phi^{(\infty)}\simeq (e^{-K_- s}-e^{-2K_- s})/(1-e^{-2K_- s})$
while $\phi^{(0)}\simeq (e^{-K_- s/2}-e^{-K_- s})/[2(1-e^{-K_- s})]$. Hence, the condition $\phi^{(\infty)}>\phi^{(0)}$
for $\phi$ to increase with $\nu$ leads to $2y^2/(1+y^2)>y/(1+y)$, where $y=e^{-K_- s/2}$. Therefore,
$\phi$ is an increasing function of $\nu/s$ when $y^2+2y-1>0$, i.e. if $y=e^{-K_- s/2}>\sqrt{2}-1$, while 
$\phi$ decreases with $\nu$ if $e^{-K_- s/2}<\sqrt{2}-1$. The critical selection pressure is thus defined by
$e^{-K_- s_c/2}=\sqrt{2}-1$. For $(K_+,K_-,s)=(450,50,0.02)$, we find $s_c\approx 0.035$. Hence, $s=0.02<s_c$ and $s=0.07>s_c$. Therefore,
 $\phi$  increases with $\nu$ when $s=0.02$, and it decreases with $\nu$ when $s=0.07$, as reported in Figs. 2(a) and \ref{fig:phi-supp}. 
\vspace{0.25cm}
\begin{figure}[h]
\begin{center}
\includegraphics[width=0.6\linewidth]{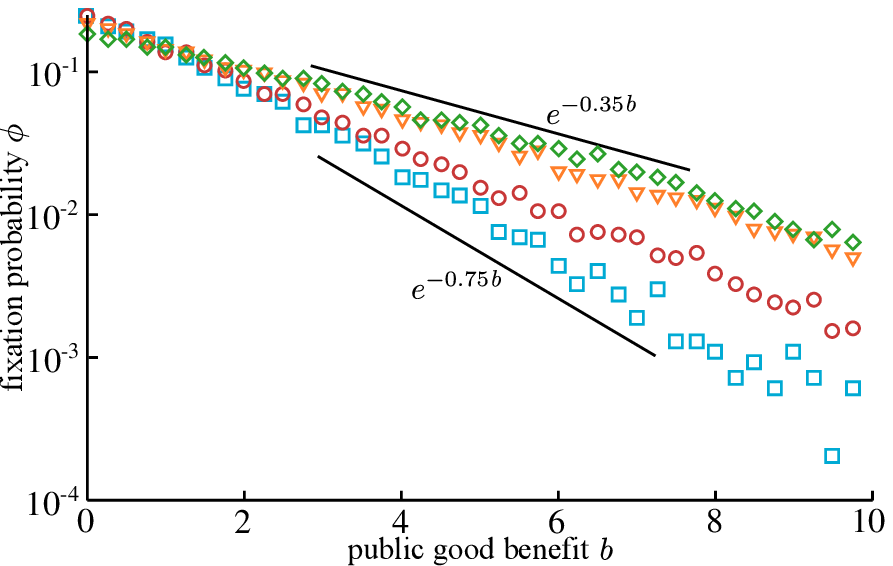}
\end{center}
\caption{{\it (Color online)}. 
$\phi$ as function of $b$ for $\nu=(0.002,0.02,0.2,2)$ (top to bottom)
and $(s,K_{+},K_-)=(0.025,450,50)$ in log scale. Straight lines show 
$0.3e^{-0.35b}$ and $0.175e^{-0.75b}$ as eyeguides.
} 
\label{fig:phib}
\end{figure}
\\
Finally, we note that while  (7) is useful to obtain an approximation of $\phi$
and its dependence on $\nu$ and $s$, it is unable to capture its dependence
on the public good parameter $b>0$. However, we know that 
the typical population size increases with $b$ when $x\approx 1$ and $S$ is close to fixation, and therefore the intensity of the demographic 
fluctuations is reduced by increasing $b$. Based on the properties of the Moran process, we thus expect  $\phi$ to  decay 
exponentially with $b$~\cite{KEM2}, which is confirmed by Fig.~\ref{fig:phib}.

\vspace{0.5cm}

\section{3. Mean fixation time}
We have also investigated the mean fixation time $T(x_0)$, which is the unconditional mean 
time  until the  fixation  of either species $S$ or $F$ starting from a initial 
 fraction $x_0$ of individuals of type $S$ in the population.
\begin{figure}[h]
\begin{center}
\includegraphics{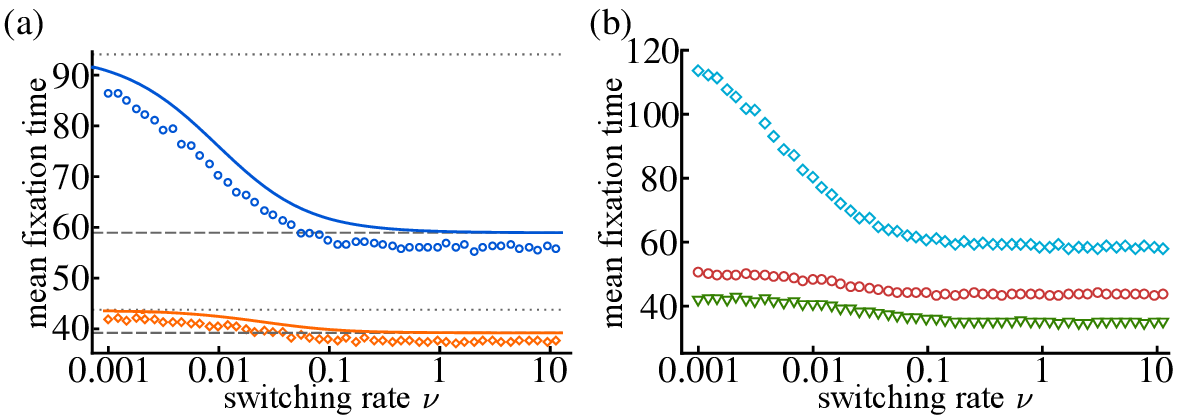}
\end{center}
\caption{{\it (Color online)}. (a)
$ T(x_0)=T $ vs. $\nu$ in the pure competition case $b=0$ with  $s=0.02$ ($\circ$, blue/black) and $s=0.07$ ($\diamond$, orange/gray).
Symbols are simulation results for $T$, solid lines are from (\ref{MFT1}), dashed and dotted lines show
 $T|_{{\cal K}}$ and $(T|_{K_-}+T|_{K_+})/2$, respectively.
(b)  $T$ vs. $\nu$ in the public good scenario 
with $b=0.2$ ($\diamond$, blue/gray for $s=0.01 $; $\circ$, red/black for $ s=0.05 $) and $b=2$, $ s=0.05 $ ($ \triangledown $, 
green/dark gray). $(K_+,K_-,x_0)=(450,50,1/2)$ in both panels. 
}
\label{fig:mft}
\end{figure}

\subsection{3.1 Mean fixation time when $b=0$}
In the case $b=0$, $N$ evolves independently of which species has fixated the population, see Videos 1-3~\cite{movies}. This allows us to
proceed just as we did with (6) for the fixation probability, and estimate the mean fixation time by $T_{\nu/s}$.
This quantity is obtained by averaging the unconditional mean fixation time $T(x_0)|_{N}$~\cite{Ewens,Cremer09} obtained in 
a Moran process for a population of constant size $ N $
over $ p^*_{\nu/s}(N) $ given by (5) with a rescaled switching rate $\nu \to \nu/s$. This yields 
\begin{eqnarray}
 \label{MFT1}
 T_{\nu/s}=\int_{K_-}^{K_+} T(x_0)|_{N}~p_{\nu/s}^*(N)~dN\,,
\end{eqnarray}
where $T(x_0)|_{N} \sim {\cal O}(1/s)$ when $e^{1/s}\gg K_-$.
As Figure \ref{fig:mft}(a) shows, this expression agrees well with the leading contribution
 $T(x_0)\simeq T_{\nu/s}\sim  {\cal O}(1/s)$ when $x_0$ is well separated from the absorbing boundaries.
 The scaling of the mean fixation time in the presence of EN is therefore the same as
 $T(x_0)|_{N}=T|_{N}$~\cite{Ewens,Cremer09}. The main effect of the EN is to affect the subleading prefactor of $T$~\cite{KEM2}:
 as shown in Fig.~\ref{fig:mft}~(a) and captured by (\ref{MFT1}), the mean fixation time
 decreases when $\nu$ increases. This stems from the fact that $\langle N \rangle^*$ 
 decreases with $\nu$, see Fig.~2(b). In the case of pure resource competition, our theory is therefore able to correctly predict 
that the mean fixation time always scales as $1/s$ but is shortened when the switching rate is increased.

\subsection{3.2 Mean fixation time when $b>0$}
In the public good scenario ($ b>0 $), the mean fixation time still scales as $T(x_0)\sim  {\cal O}(1/s)$
and decreases with the environmental switching rate $ \nu $, as shown in \ref{fig:mft}(b). This is because the average
population size also decreases with $ \nu $ (see Fig. 3). In this case, 
however the fixation of the $S$ type happens in larger populations (and, hence, after longer times) than the fixation of $F$, see 
Videos 6-7~\cite{movies}. 
As a result, to accurately compute $ T(x_0)$, it is necessary to determine the two conditional mean fixation times 
(which are equal only when $b=0$)~\cite{KEM2}. Clearly, this cannot be achieved by assuming a timescale separation between $ N $ and $ x $, and is beyond the reach of 
our effective theory. More precisely, it is necessary to generalize the effective theory in order to compute the mean fixation times when $b>0$~\cite{KEM2}.

\section{4. Population size quasi-stationary distribution: additional discussion and results}
In this section, we provide additional discussion and results about the population size distribution 
after the occurrence of fixation. An important common feature of the $b=0$ and $b>0$ scenarios is that 
long-time population size distribution is well described by $p_{\nu}^*$ (5) when $b=0$,
and by combining the conditional PDFs $p_{\nu}^*$ and $p_{\nu,b}^*$ (\ref{pstarq}) with $\phi$ when $b>0$, 
as explained in the main text.

\subsection{4.1 Noise-induced transitions}
The quasi-stationary  population size distributions are thus characterized by different regimes in which they are unimodal,
bimodal, or even multimodal, see Figs. 3, 4 and \ref{fig:n-dist-conditional}. The transitions between these various 
regimes  are called ``noise-induced transitions'' because they are solely caused by the environmental noise~\cite{HL06,Bena06}.
In fact, if the carrying capacity in (\ref{SDEaux}) was oscillating periodically (deterministically), 
the corresponding PDF would {\it always} be bimodal: the transition to the unimodal regime is only possible for randomly fluctuating $ K $~\cite{Bena06}.

\subsection{4.2 Simulation and prediction of the population size steady state distribution}
To assess the theoretical predictions for the long-time population size distribution inferred from
(5) and (\ref{pstarq}), we have generated $10^5$ replicas that we let run until $99\%$
of them reached fixation. The outcome has then been binned to generate the histograms shown 
as solid lines in Figs.~3, 4(c,d) and in Fig.\ref{fig:n-dist-conditional}.
\\
In the pure competition case ($ b=0 $), see Fig.~3, these simulation results are compared with 
$ p^*_\nu(N) $ (5) multiplied by the number of replicas. (In this case, $N$
evolves independently of $x$, therefore it is not necessary to wait until $ 99\% $ of fixation has occurred, see 
Videos 4-5~\cite{movies}.
We have  proceeded in this way for consistency with the case $ b>0 $).
\\ 
In the public good scenario ($b>0$), see Figs. 4(c,d) and  \ref{fig:n-dist-conditional}, 
we have waited until fixation had occurred in almost all replicas ($99\%$ of them) to collect 
the data to build the histograms that correctly reflect the quasi-stationary state distributions of the 
population size (now depending on $x$), see Videos 8-10~\cite{movies}.
Via our effective theory, we have computed the fixation probability of the strain $S$ and $F$. Multiplying these 
values by $ 10^5 $ (number of samples), we have obtained the expected number of replicas to fixate to $S$
and to $F$. By multiplying these numbers by $p_{\nu}^*$ (5) and $p_{\nu,b}^*$ (\ref{pstarq}) we obtain
the histograms associated with the conditional probability distributions (unconditioned of $\xi=\pm 1$).
These are shown by dotted lines in Fig. \ref{fig:n-dist-conditional}
and their sum gives the histogram of the marginal distribution 
(orange dashed lines in Fig \ref{fig:n-dist-conditional}), which can be directly 
compared with the histogram from the simulations.

\begin{figure}[htb]
\begin{center}
\includegraphics{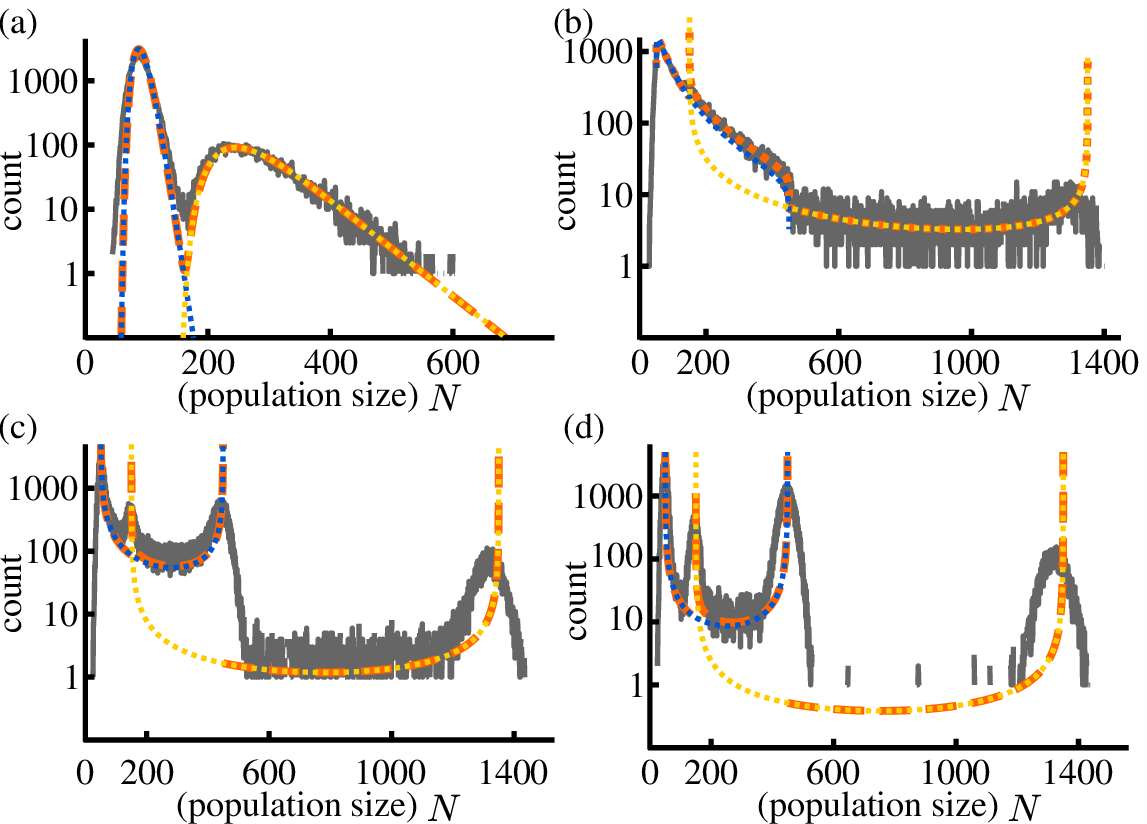}
\caption{ {\it (Color online)}. Long-time population
size distributions for  $\nu=20$ (a), $\nu=1.2$ (b), $\nu=0.2$ (c), and $\nu=0.02$ (d)
with $(K_+,K_-,x_0,s,b)=(450,50,0.5,0.02,2)$
similar to Fig.~4(c,d) but now showing also the results obtained from the 
$S$-conditional (dotted, yellow/light gray) and $F$-conditional 
(dotted,  blue/dark gray) PDFs. The histogram of the marginal 
PDF (dashed) is the sum of the $S/F$-conditional histograms
weighted by $\phi_q$ (7), see text.}
 \label{fig:n-dist-conditional}
 \end{center}
\end{figure}

\subsection{4.3 Long-time population size distribution in the public good scenario ($b>0$)}

To understand the properties of the quasi-stationary marginal population size distribution when $b>0$, 
it is useful to notice that when $S$ fixates ($x=1$), the relevant conditional PDF (unconditioned of $\xi=\pm 1$) is 
$p_{\nu,b}^*$ which is unimodal and peaked at $N=(1+b){\cal K}$
when $\nu>1+b$, while it is bimodal with peaks at $N=(1+b)K_{\pm}$
if $\nu<1+b$. Similarly, $p_{\nu}^*$  is the  
PDF conditioned to fixation of $F$ (but unconditioned of $\xi=\pm 1$): it  is  unimodal and peaked at $N\approx{\cal K}$
if $\nu>1$, whereas it is bimodal with  peaks at $N\approx K_{\pm}$
when $\nu<1$. The sum of the conditional  PDFs weighted by $\phi_q$ 
yields the marginal PDF (unconditioned of $\xi=\pm 1$ and of whether $S$ or $F$ fixates) that, depending on $\nu$ and $b$, is either bimodal or 
multimodal. Therefore, as shown in Figs.~4(c,d) and \ref{fig:n-dist-conditional} as well as in Videos 8-10~\cite{movies},
the marginal quasi-stationary  population size distribution is characterized by 
\begin{enumerate}
 \item[-] two  peaks  at about $N={\cal K}$ and $N=(1+b){\cal K}$  when $\nu>1+p$, see Video 8.
 \item[-] 
  three peaks located about $N={\cal K}$ and $N=(1+b)K_{\pm}$ when $1<\nu<1+p$, see Video 9.
 \item[-] four peaks located around $N=(1+b)K_{\pm}$ and  $N=K_{\pm}$ when $\nu<1$, see Video 10. 
\end{enumerate}
The peaks  at $N=(1+b)K_{\pm}$ and $N=(1+b){\cal K}$ stem from the fixation of 
$S$ and thus are less marked than those at $N\approx K_{\pm}$ and $N\approx {\cal K}$ which result from the
more likely fixation of $F$.

\subsection{4.4 Figure 4(c,~d) revisited}
In Fig.~4(c,d), we report the histograms of the stationary marginal population distribution at $\nu=20$ and
$\nu=0.02$ with $b=2$. For the sake of  completeness, in Fig.~\ref{fig:n-dist-conditional} we also consider  the intermediate
switching rates $\nu=1.2$  and $\nu=0.2$, and show the conditional PDFs $p_{\nu}^*$ and $p_{\nu,b}^*$.
The marginal PDF is obtained as the sum of $p_{\nu}^*$ and $p_{\nu,b}^*$
weighted by $\phi_{q}$ and $1-\phi_{q}$ given by (7).

\subsection{4.5 Deviations from the PDF predictions}
We have seen that coupled internal and environmental noise greatly influences the population fixation probability (aptly 
described by Eqs.(6) and (7)), and therefore significantly influences the population internal composition 
(evolutionary dynamics), and in turn also its ecological dynamics when $b>0$ (internal and ecological dynamics being 
then explicitly coupled). We have also seen that once fixation has occurred, the 
population size quasi-stationary distribution is well described by the stationary (conditional) PDFs (5) and  (\ref{pstarq})  of underlying PDMP
that are able to predict when the long-time population size distributions are unimodal, 
bimodal or multimodal and the location of the peaks, as shown by 
 Figs.~3, 4 and \ref{fig:n-dist-conditional}.\\
 However, Eqs.~(5) and (\ref{pstarq}) ignore the effects of demographic fluctuations 
 on the population size distribution. In fact,
 demographic fluctuations are responsible for the population size quasi-stationary  distributions
obtained from the simulations not to be strictly confined within the support of the PDFs (5) and 
(\ref{pstarq}), especially at low $\nu$, as can be seen in Figs.~3, 4 and \ref{fig:n-dist-conditional}. As clearly visible in the 
supporting Videos~\cite{movies}, 
these deviations appear because, due to demographic noise,   
the population fluctuates around the fixed points $ N=K_{\pm} $ and $ N=(1+b)K_{\pm} $, see Video 10.
The small deviations from the PDMP predictions have limited influence on quantity such as the average population size
$\langle N \rangle ^*$, see main text, and their intensity depends on the values of $K_\pm$ (high values of $K_\pm$
typically yield broader peaks)~\cite{KEM2}.

\section{5. Supporting videos}
The dynamics of the models and our findings are illustrated by a series of videos available 
electronically~\cite{movies}. 
\subsection{5.1 Videos 1-5: $b=0$}
Videos 1-5 illustrate the population dynamics in the pure resource competition scenario for the parameters
 $(s,K_+,K_-,x_0)=(0.02, 450,50,0.5)$ and different switching rates. (In all videos, various initial values of the population size, $N(0)$,  
have been considered, but, after a brief transient, these have no influence on the results).
\begin{itemize}
\item Video 1 shows the sample paths $N(t)$ (left) and $x(t)$ (right) of five replicas for $\nu=20$. We
 clearly notice a timescale separation: the population size quickly starts to endlessly fluctuate
 about $N\approx {\cal K}=90$ while $x(t)$  evolves much more slowly, with fixation occurring in time $t\sim {\cal O}(1/s)$.
\item Video 2 shows similar paths for $\nu=0.01$ (and a sped-up animation). We again see the timescale separation between
 $N(t)$ and $x(t)$. However, in the long run $N(t)$  endlessly jumps between $N\approx K_-$ and $N\approx K_+$.
Moreover, the video shows how the behavior of the population size is unaffected by changes in $x$: $N$ relaxes at a 
faster timescale 
and maintains the same behavior also after fixation (of either species).
\item Video 3: $N(t)$ and $x(t)$ sample paths as in Videos 1 and 2 but for very slow switching rate $\nu=0.0001\ll s=0.02$.
In all but one replicas, the population evolves subject to the carrying capacity $K_-$ or $K_+$, randomly
allocated initially with same probability, without experiencing any switches and $N(t)$ fluctuates about $K_+$ or $K_-$
In only one realization,  after a long time (at $t\approx 750$), the  carrying capacity switches and the 
population jumps from $K_+$ to $K_-$. The video also illustrates that $S$ fixation is more likely when the 
population is subject to $K=K_-$ than to $K=K_+$: both the purple and pink
samples ending at $x=1$ correspond to a population of size $N(t)\approx K_-$.
\item Video 4 
shows the histograms of the population size (left) and  of the fraction of $S$ individuals (right) 
for a slow-switching environment ($\nu=0.2$). 
We notice that the population size distribution readily attains a right-tailed, bimodal shape
with peaks  about 
$N=K_{\pm}$, and  is independent of the distribution of $x$ (internal dynamics).
On the other hand, the histogram of $x$ evolves slowly and is eventually characterized by asymmetric peaks at $x=0$ and $x=1$ 
corresponding to the fixation probability of $F$ and $S$, respectively.
\item Video 5: as in Video 3, but for a fast-switching environment ($\nu=20$).
 The population size histogram rapidly becomes bell-shaped and centered about 
$N={\cal K}$. It reaches this form much before  fixation typically occurs, and is independent of the distribution of 
$x$ (internal dynamics). 
 The histogram of $x$ has the same properties as in Video 3.
\end{itemize}
 
\subsection{5.2 Videos 6-10: $b>0$} 
 Movies 6-8 illustrate the internal and ecological dynamics in the public good scenario, $b>0$, for the parameters
 $(s,K_+,K_-,b,x_0)=(0.02, 450,50,2,0.5)$ and different switching rates.
  In this scenario, the fast $ N $ dynamics is enslaved to the slower evolution of $ x $. 
  The population size distribution is characterized by peaks that slowly emerge 
as  occurrences of $S$ and $F$ fixation accumulate (right panels).
\begin{itemize}
\item Video 6 shows sample paths of $N$ and $x$  for five realizations with $\nu=20$, as in Video 1. 
The population size and composition are  correlated: the population size attains large values when
 $x$ dwells about $1$, while $N$ is much smaller when $x\ll 1$
 (for example, the green replica is almost always 
 larger than the purple one). As the species fixate, the sample paths for $N$ separate 
 into two distinct sets: those associated with the fixation of $S$ ($x=1$)
 fluctuate about $N\approx {\cal K}=90$, while the paths associated to $x=0$ (fixation of $F$)
 fluctuate around $N\approx (1+b){\cal K}=270$.
\item Video 7 shows similar sample paths for $\nu=2$. In addition to showing
the correlation between $N$ and $x$, the video illustrates how
populations with a high fraction of $S$ ($x\approx 1$)
experience random switching with an effectively reduced switching rate. For example, in
the purple sample paths, which readily attains $x\approx 1$,
$N$ evolves by large abrupt jumps, in agreement with the properties of the $S$-conditional PDF
$p^*_{\nu,2}$, see (\ref{pstarq}). 
\item Video 8 shows the histograms of $N$ and $x$ for fast switching ($\nu=20$). The histogram of the population size  (left) 
has first a right-tailed bell shape. As fixation occurrences build up, the distribution gradually splits into
asymmetric peaks about ${\cal K}=90$ and $(1+b){\cal K}=270$. The histogram of $x$ 
is characterized by slowly-developing asymmetric peaks at $x=0$ and $x=1$.
\item Video 9 shows the histograms of $N$ and $x$ for intermediate switching ($\nu=1.2$). Similarly to Video 7, 
the histogram of $N$ changes from having first a right-tailed bell shape to its eventual quasi-stationary form. In this case, the 
quasi-stationary  state is characterized by three asymmetric peaks, located at about ${\cal K}=90$ and $(1+b)K_-=150$, and 
about $(1+b)K_+=1350$, that slowly develop as  fixation occurrences pile up (right panel).
\item Video 10 shows the histograms of $N$ and $x$ for slow switching $\nu=0.2$.
Initially, the histogram of $N$ develops as in Videos 7 and 8, but now the quasi-stationary  state is characterized by four
slowly-developing asymmetric peaks, located at about $K_-=50$,  $(1+b)K_-=150$, and about $K_+=450$  
and $(1+b)K_+=1350$.
\end{itemize}
\newpage
\end{widetext}


\begin{thebibliography}{99}
\bibitem{footnote}
 For instance,  for $(K_{+}, K_-, s, b, x_0)=(450,50,0.025,2,0.5)$ we have found $q=1.1543$.
\end{thebibliography}
\end{document}